\documentclass[12pt]{article}

\usepackage[a4paper, margin=1in]{geometry}
\usepackage{setspace}
\usepackage{amsmath, amssymb, amsthm}
\usepackage{booktabs}
\usepackage{enumitem}
\usepackage{microtype}
\usepackage{parskip}
\usepackage{titlesec}
\usepackage{xcolor}
\usepackage{array}
\usepackage{multirow}
\usepackage{caption}
\usepackage{float}
\usepackage[section]{placeins}
\usepackage{graphicx}
\usepackage[utf8]{inputenc}
\usepackage[authoryear,round]{natbib}
\usepackage{bm}
\usepackage{algorithm}
\usepackage{algpseudocode}

\definecolor{darkblue}{rgb}{0.05,0.10,0.40}
\usepackage[colorlinks=true,linkcolor=darkblue,citecolor=darkblue,urlcolor=darkblue]{hyperref}

\titleformat{\section}{\large\bfseries}{\thesection}{0.6em}{}
\titleformat{\subsection}{\normalsize\bfseries}{\thesubsection}{0.5em}{}
\titleformat{\subsubsection}{\normalsize\itshape}{\thesubsubsection}{0.5em}{}
\titlespacing{\section}{0pt}{12pt}{4pt}
\titlespacing{\subsection}{0pt}{8pt}{2pt}
\titlespacing{\paragraph}{0pt}{6pt}{0.5em}

\newtheorem{proposition}{Proposition}

\newcommand{\bX}{\mathbf{X}}
\newcommand{\bA}{\mathbf{A}}
\newcommand{\bY}{\mathbf{Y}}
\newcommand{\bZ}{\mathbf{Z}}
\newcommand{\bQ}{\mathbf{Q}}
\newcommand{\bM}{\mathbf{M}}
\newcommand{\bK}{\mathbf{K}}
\newcommand{\bw}{\mathbf{w}}
\newcommand{\bp}{\mathbf{p}}
\newcommand{\bq}{\mathbf{q}}
\newcommand{\ba}{\mathbf{a}}
\newcommand{\bz}{\mathbf{z}}
\newcommand{\bu}{\mathbf{u}}

\newcommand{\bv}{\mathbf{v}}

\newcommand{\beps}{\boldsymbol{\varepsilon}}
\newcommand{\balpha}{\boldsymbol{\alpha}}
\newcommand{\bbeta}{\boldsymbol{\beta}}

\newcommand{\SX}{\mathbf{V}_X}
\newcommand{\SZ}{\mathbf{V}}

\newcommand{\cor}{\operatorname{cor}}
\newcommand{\argmax}{\operatorname{arg\,max}}
\newcommand{\argmin}{\operatorname{arg\,min}}

\newcommand{\calO}{\mathcal{O}}

\begin{document}
\setlength{\abovedisplayskip}{6pt plus 2pt minus 2pt}
\setlength{\belowdisplayskip}{6pt plus 2pt minus 2pt}
\setlength{\abovedisplayshortskip}{3pt plus 2pt}
\setlength{\belowdisplayshortskip}{6pt plus 2pt minus 2pt}
\begin{singlespacing}

\vspace*{10pt}
\begin{center}
  {\LARGE\bfseries Learning the Optimal Composite Mediator:\\[6pt]
    Closed-Form Solution and Inference}\\[14pt]
  {\large Zihuai He}\\[3pt]
  {\small\href{https://orcid.org/0000-0002-8220-4183}{\texttt{orcid.org/0000-0002-8220-4183}}}\\[5pt]
  {\normalsize
    Quantitative Sciences Unit, Department of Biomedical Data Science,\\
    and Department of Neurology \& Neurological Sciences,\\
    Stanford University\\[3pt]
    \texttt{zihuai@stanford.edu}
  }\\[10pt]
  {\small\textit{Running head:}\/ Optimal Composite Mediator}
\end{center}

\vspace{14pt}

\begin{center}{\small\bfseries Abstract}\end{center}
\vspace{2pt}
{\small\noindent
Understanding how an exposure transmits its effect through high-dimensional
intermediaries is a central problem in observational research.
We study the problem of finding a composite mediator that maximises the
indirect effect of an exposure on an outcome in a linear structural equation model.
Although the objective is non-convex in the weight vector, a geometric
argument yields a closed-form global solution: the optimal weight bisects
the angle between two computable path vectors in a weighted inner product
space, recovered via two linear solves.
The resulting algorithm, \textbf{MaxIE}, runs at the same cost as ordinary
least squares---orders of magnitude lower than numerical optimisation---with
a dual formulation for settings where mediators outnumber observations.
The same path vectors yield a test for the global null that
no composite mediator exists, with $t(p-1)$ in the classical and $t(n-2)$ in the dual regime.
Power is characterised analytically as a function of the population path angle;
simulations confirm size control and the power characterisation.
Applied to a UK Biobank proteomics dataset ($n=38{,}383$, $p=2{,}916$),
the method rejects the global null ($p\text{-value} = 6.4\times10^{-9}$)
and identifies the optimal proteomic composite mediating age's effect on dementia.
}

\vspace{8pt}


\end{singlespacing}
\newpage

\section{Introduction}
\label{sec:introduction}

Many contemporary studies involve a treatment or exposure $\bA$ whose effect
on an outcome $\bY$ is transmitted through a large set of candidate mediators
$\bX \in \mathbb{R}^{n \times p}$, such as gene expression profiles or
protein abundances. Mediation analysis asks through which intermediaries this
effect operates---knowledge that can point to therapeutic targets, modifiable
risk factors, or mechanisms linking environmental exposures to disease.

The mediated effect of a treatment often reflects the collective activity of
many molecular features rather than any single one: genes act in regulatory
pathways, proteins function within signalling complexes, and metabolites
participate in coordinated biochemical reactions. A composite score
$\bM = \bX\bw$ is therefore a natural target---one that aggregates the joint
mediation signal into a single interpretable mediator and captures structure
that single-feature analyses miss. A particularly pressing instance arises in
aging research, where large-scale proteomics platforms now enable construction
of \emph{proteomic aging clocks}---composite protein scores that estimate
biological age and predict a wide range of diseases including Alzheimer's,
heart failure, and cancer~\citep{oh2023,oh2025,argentieri2024,tian2025,wang2025}.
These composites are increasingly used to track biological aging and stratify
disease risk, and are natural candidates for monitoring the effects of
anti-aging interventions~\citep{oh2023,oh2025,argentieri2024,tian2025,wang2025}.
Yet existing clocks are constructed by regressing chronological age on proteins
alone, without reference to any disease outcome, maximising age tracking at the
expense of disease relevance.

The mediation framing resolves this tension directly: chronological age
remodels the proteome, and those proteomic changes are the mechanism
through which age increases disease risk~\citep{oh2023,tian2025}.
The composite mediator of interest is a protein score that
\emph{simultaneously} tracks biological aging and transmits that signal to
disease risk---yet this joint optimisation has not previously been pursued
as a direct problem, and neither single-criterion approach guarantees that
the resulting composite jointly optimises both properties.
\paragraph{The mediation objective.}
We model mediation through three structural equations on centred
variables~\citep{baron1986}:
\begin{equation}
  \bM = \alpha\, \bA + \beps_1, \qquad
  \bY = \gamma\, \bA + \beta\, \bM + \beps_2, \qquad
  \bY = \tau\, \bA + \beps_3,
  \label{eq:bk}
\end{equation}
where $\alpha$, $\beta$, $\gamma$, $\tau$ are path coefficients (treatment$\to$mediator,
direct, mediator$\to$outcome, total) and $\beps_1,\beps_2,\beps_3$ are mean-zero errors.
We seek the weight vector $\bw^*$ that maximises the population indirect
effect for the composite mediator $\bM = \bX\bw$:
\begin{equation}
  \bw^*_{+} \;=\; \argmax_{\bw \in \mathbb{R}^p} \;\alpha(\bw)\,\beta(\bw),
  \label{eq:objective}
\end{equation}
estimated in practice by $\argmax_\bw \hat\alpha(\bw)\hat\beta(\bw)$.
This finds the \emph{concordant} composite mediator: the direction along which
treatment increases the mediator \emph{and} the mediator increases the outcome
(both path coefficients share the same sign).
A complementary quantity is the \emph{suppression} composite mediator,
\begin{equation}
  \bw^*_{-} \;=\; \argmin_{\bw \in \mathbb{R}^p} \;\alpha(\bw)\,\beta(\bw),
  \label{eq:objective_supp}
\end{equation}
which captures directions where treatment and outcome signals pull in
opposite directions ($\alpha(\bw)\beta(\bw) < 0$).
Since flipping the sign of $\bw$ changes the signs of both $\alpha(\bw)$
and $\beta(\bw)$ simultaneously, their product is unchanged; we orient
each solution so that
$\hat\alpha(\bw^*_{\pm}) \geq 0$, after which the sign of $\hat\beta$ identifies
concordant ($\hat\beta>0$) versus suppression ($\hat\beta<0$) mediation.
Running both recovers a complete picture of how treatment effects are
transmitted through the mediator panel.
The remainder of the paper focuses on the concordant estimator $\bw^*_+$
without loss of generality: as shown in Section~\ref{sec:setup},
$\bw^*_-$ is obtained by negating the sufficient statistic $\bz$, so all
theoretical results and algorithms apply to both problems by a trivial sign
change at no additional cost.

\subsection*{Related Work}
\paragraph{Causal mediation analysis.}
The modern causal framework for mediation, developed by
Robins \& Greenland~\citep{robins1992} and Pearl~\citep{pearl2001} and extended by
Imai et al.~\citep{imai2010} and VanderWeele~\citep{vanderweele2015}, defines the natural indirect effect (NIE)
and natural direct effect (NDE) in terms of potential outcomes, with
identification resting on sequential ignorability and the cross-world
consistency assumption.
Under a linear structural equation model these reduce to the
path-coefficient product $\alpha\beta$~\citep{pearl2001}, which is the
estimand we maximise.
Semiparametric efficiency bounds for the NIE are derived in
Tchetgen Tchetgen \& Shpitser~\citep{tchetgen2012}; our contribution operates
within the parametric linear model and solves the previously unaddressed
problem of optimally choosing the composite mediator within this class.
\paragraph{High-dimensional composite mediator methods.}
Principal component mediation analysis (PCMA,~\citep{pcma}) and its sparse
variant SPCMA~\citep{spcma} construct composite scores as principal components
of the treatment-residualised mediators.  These methods are designed for
dimension reduction and screening: the PCA criterion maximises residual
variance in $\bX$, which is independent of both $\bA$ and $\bY$.
They do not optimise any mediation criterion and are therefore solving a
different problem from ours; a leading principal component need not align with
either mediation path.
The most direct predecessor to MaxIE is the Directions of Mediation
(DM,~\citep{chen2018dm}), which finds a composite mediator by maximising the
joint LSEM log-likelihood via an iterative eigenvalue algorithm.  Both DM and
MaxIE seek a single weight vector capturing the mediation pathway, but DM uses
a likelihood surrogate and requires iterative optimisation, whereas MaxIE
directly maximises $\hat\alpha(\bw)\hat\beta(\bw)$ in closed form.
Moreover, DM provides no inferential framework for the global null that no
composite mediator of either sign exists ($H_0$:~\eqref{eq:H0}); the cosine test derived here fills this gap.
\paragraph{Screening and selection methods.}
A large literature screens individual mediators via penalisation:
HIMA~\citep{zhang2016}, HDMA~\citep{hdma}, Zhou et al.~\citep{zhou2020}, and
Pathway LASSO~\citep{pathwaylasso} recover active individual mediators, but
any composite score they produce is a byproduct of regularisation rather than
the result of directly optimising a composite indirect-effect criterion.
\paragraph{Our contributions.}
\textit{(i) Closed-form optimal composite mediator, primal and dual.}
Jointly optimising $\alpha(\bw)\beta(\bw)$ as a direct problem has not
previously been attempted.
In the \textbf{primal} regime ($n > p$), the objective factorises into a
path-strength term and an alignment term; the optimal $\bw$ lies in the span
of two computable path vectors and is recovered by two linear solves plus a
bisector evaluation at cost $\mathcal{O}(np + p^3/3)$
(Section~\ref{sec:indirect}).
In the \textbf{dual} regime ($p \geq n$), the problem reduces to an
$n \times n$ kernel eigendecomposition via $\bK_Z = \bZ\bZ^\top$, recovering
the optimal weight in the row space of $\bX$ at cost $\mathcal{O}(n^2 p + n^3)$---the
only algorithm for the $p \geq n$ setting.
Both concordant and suppression solutions are returned by a single call at no
additional cost.

\textit{(ii) Global test for composite mediation, primal and dual.}
We derive an asymptotic test for $H_0: \max_\bw |\alpha(\bw)\beta(\bw)|=0$
based on $\cos\hat\varphi$, the cosine of the angle between the two path vectors
(Section~\ref{sec:inference}).
\textbf{Primal}: $T \sim t(p-1)$ asymptotically as $n\to\infty$, $p$ fixed;
exact for any $n>p$ when the $\beta$-path is null, asymptotic when the
$\alpha$-path is null.
\textbf{Dual}: $T \sim t(n-2)$ asymptotically as $p\to\infty$, $n$ fixed;
exact for any $p>n$ when the $\beta$-path is null, asymptotic when the
$\alpha$-path is null.
Power is characterised analytically in both regimes; the dual has a saturation
ceiling $\pi_\infty(\varphi_0, n) < 1$.
Both tests are free by-products of the solver.

\section{Problem Setup}
\label{sec:setup}

We work within the linear structural equation model~\eqref{eq:bk}.
In the potential-outcomes language of Robins \& Greenland~\citep{robins1992} and Pearl~\citep{pearl2001},
let $Y(a,m)$ denote the potential outcome under treatment $a$ and mediator
value $m$, and $M(a)$ the potential mediator value under treatment $a$.
The natural indirect effect at treatment level $a$ is
$\mathrm{NIE}(a) = E[Y(a, M(1)) - Y(a, M(0))]$.
Under the linear structural model~\eqref{eq:bk}, this reduces to
$\mathrm{NIE}(a) = \alpha\beta$ for any $a$, so the product $\alpha(\bw)\beta(\bw)$
is the NIE for composite mediator $\bM = \bX\bw$.
The following assumptions are maintained throughout.

\begin{itemize}[leftmargin=2em,itemsep=2pt]
\item[\textbf{A1}] \textit{(Consistency)} The observed outcome satisfies
  $Y_i = Y_i(A_i, M_i(A_i))$ and the observed mediator satisfies
  $M_i = M_i(A_i)$ for each $i$.
\item[\textbf{A2}] \textit{(Sequential ignorability)} (a)~$\{Y(a,m), M(a')\}
  \perp\!\!\!\perp A \mid \bX$; (b)~$Y(a,m) \perp\!\!\!\perp M \mid A, \bX$,
  for all $a, a', m$ in their respective supports.
\item[\textbf{A3}] \textit{(Linear structural model)} The structural
  equations~\eqref{eq:bk} hold with mean-zero, finite-variance errors
  $\beps_1, \beps_2, \beps_3$ independent of $(\bA, \bX)$.
\item[\textbf{A4}] \textit{(Full rank)} $\bX$ has full column rank and
  $\Sigma_Z = n^{-1}E[\bZ^\top\bZ]$ is positive definite.
\end{itemize}

Under A1--A3, $\alpha(\bw)\beta(\bw)$ equals the population NIE for composite
mediator $\bX\bw$~\citep{pearl2001,vanderweele2015}.
Our contribution is tractable optimisation and inference within this parametric
model; relaxing A3 to semiparametric models is an important open direction
(see Section~\ref{sec:conclusion}).

Let $\bX \in \mathbb{R}^{n \times p}$ be the matrix of candidate mediators,
$\bA \in \mathbb{R}^n$ the treatment vector, and $\bY \in \mathbb{R}^n$ the
outcome vector, all centred, with $n \geq p$ and $\bX$ full column rank.
We form a composite mediator $\bM = \bX\bw$ and seek
\begin{equation}
  \bw^*_{+} \;=\; \argmax_{\bw \in \mathbb{R}^p}\; h(\bw), \qquad
  h(\bw) \;=\; \hat\alpha(\bw)\,\hat\beta(\bw),
  \label{eq:hobj_def}
\end{equation}
where $\hat\alpha(\bw)$ and $\hat\beta(\bw)$ are the OLS path coefficients
from the structural regressions~\eqref{eq:bk}.
Applying OLS with mediator $\bM = \bX\bw$ gives
\begin{equation}
  \hat\alpha(\bw) \;=\; \frac{\bA^\top(\bX\bw)}{\|\bA\|^2}, \qquad
  \hat\beta(\bw) \;=\; \frac{(\bZ\bw)^\top\bY}{\|\bZ\bw\|^2},
  \label{eq:pathcoeffs}
\end{equation}
where $\bQ_A = I_n - \bA\bA^\top/\|\bA\|^2$ is the orthogonal projection
onto the complement of $\bA$, and $\bZ = \bQ_A\bX$ is the matrix of
$\bA$-residualised mediators.
To reparametrise $h(\bw)$ in terms of $\bw$ alone, we define three
$p$-dimensional sufficient statistics:
\begin{align}
  \ba  &\;=\; \bX^\top\bA, \qquad
  \bz  \;=\; \bX^\top\bY - \frac{\bA^\top\bY}{\|\bA\|^2}\,\ba, \label{eq:az} \\
  \SZ  &\;=\; \bX^\top\bX - \frac{\ba\ba^\top}{\|\bA\|^2}
         \quad\text{(the residual Gram matrix).} \label{eq:sigmas}
\end{align}
One can verify that $\SZ = \bZ^\top\bZ$ and $\bz = \bZ^\top\bY$, so these
are computable via rank-1 updates to $\bX^\top\bX$ and scalar inner products,
without forming $\bZ$ explicitly.
In terms of $(\ba, \bz, \SZ)$, the path coefficients become
$\hat\alpha(\bw) = \bw^\top\ba/\|\bA\|^2$ and
$\hat\beta(\bw) = \bw^\top\bz/(\bw^\top\SZ\bw)$, and $h(\bw)$ reparametrises as
\begin{equation}
  h(\bw) \;=\; \frac{(\bw^\top\ba)\,(\bw^\top\bz)}{\|\bA\|^2\,(\bw^\top\SZ\bw)}.
  \label{eq:hobj}
\end{equation}
Since $h(\bw)$ is scale-invariant in $\bw$, we optimise over directions.
By symmetry, $\min_{\bw}h(\bw) = \max_{\bw}h(\bw;{-}\bz)$,
where $h(\bw;{-}\bz)$ denotes $h(\bw)$ with $\bz$ replaced by $-\bz$,
so both the concordant ($\bw^*_+$) and suppression ($\bw^*_-$) problems
are solved by the same algorithm.
Because the suppression direction uses $-\bq = \SZ^{-1}(-\bz)$, already
computed, both solutions are returned by a single function call at no
additional computational cost (Section~\ref{sec:indirect}).

\section{Closed-Form Solution for $\hat\alpha(\bw)\hat\beta(\bw)$}
\label{sec:indirect}

The objective $h(\bw)$ in~\eqref{eq:hobj} is non-convex: the numerator is a
product of two linear forms in $\bw$, so gradient-based methods can converge
to local optima.  However, $h(\bw)$ is scale-invariant---replacing $\bw$ by
$\lambda\bw$ leaves it unchanged---so the search is over directions only.
Working in the appropriate inner product, this geometric view leads to a
closed-form global solution.

\subsection{Geometric Reduction}

Since $\SZ = \bZ^\top\bZ$ is positive definite, it defines an inner product
on $\mathbb{R}^p$: for any $\bu_1, \bu_2 \in \mathbb{R}^p$,
\[
  \langle \bu_1, \bu_2 \rangle_{\SZ} \;=\; \bu_1^\top\SZ\,\bu_2, \qquad
  \|\bu_1\|_{\SZ} \;=\; \sqrt{\bu_1^\top\SZ\,\bu_1},
\]
with $\bu_1 \perp \bu_2$ meaning $\langle \bu_1, \bu_2 \rangle_{\SZ} = 0$,
and the angle between $\bu_1$ and $\bu_2$ defined by
$\cos\angle(\bu_1,\bu_2) = \langle\bu_1,\bu_2\rangle_{\SZ}/
(\|\bu_1\|_{\SZ}\|\bu_2\|_{\SZ})$.
All geometry in this section is with respect to this inner product.

Define the two path vectors
\begin{equation}
  \bp \;=\; \SZ^{-1}\ba, \qquad \bq \;=\; \SZ^{-1}\bz.
  \label{eq:pq}
\end{equation}
Because $\bp = \SZ^{-1}\ba$, inner products reduce to dot products with the
sufficient statistics: $\langle\bp,\bq\rangle_{\SZ} = \ba^\top\bq$,
$\|\bp\|_{\SZ} = \sqrt{\ba^\top\bp}$, $\|\bq\|_{\SZ} = \sqrt{\bz^\top\bq}$.
Since $\SZ\bp = \ba$ and $\SZ\bq = \bz$, the numerator factors rewrite as
\[
  \bw^\top\ba \;=\; \|\bw\|_{\SZ}\|\bp\|_{\SZ}\cos\angle(\bw,\bp), \qquad
  \bw^\top\bz \;=\; \|\bw\|_{\SZ}\|\bq\|_{\SZ}\cos\angle(\bw,\bq),
\]
while the denominator is $\bw^\top\SZ\bw = \|\bw\|_{\SZ}^2$.
Substituting into~\eqref{eq:hobj}, the $\|\bw\|_{\SZ}^2$ cancels and the
objective factorises as
\begin{equation}
  h(\bw) \;=\;
    \underbrace{\frac{\|\bp\|_{\SZ}\,\|\bq\|_{\SZ}}{\|\bA\|^2}}_{\text{path strength}}
    \;\cdot\;
    \underbrace{\cos\angle(\bw,\bp)\cdot\cos\angle(\bw,\bq)}_{\text{alignment}(\bw)}.
  \label{eq:hdecomp}
\end{equation}
The path-strength factor depends only on the data, not on $\bw$.
Maximising $h(\bw)$ therefore reduces to maximising the alignment term:
given that the angle $\varphi = \angle(\bp,\bq)$ between the two path
vectors is fixed, find the direction $\bw$ that maximises
$\cos\angle(\bw,\bp)\cdot\cos\angle(\bw,\bq)$.

\begin{proposition}[Bisector optimum]
\label{prop:bisector}
Let $\varphi = \angle_{\SZ}(\bp,\bq)$ be the angle between the two path vectors.
The alignment $\cos\angle_{\SZ}(\bw,\bp)\cdot\cos\angle_{\SZ}(\bw,\bq)$ is maximised
by the direction $\bw^*$ that bisects the angle between $\bp$ and $\bq$,
i.e.\ $\angle_{\SZ}(\bw^*,\bp) = \angle_{\SZ}(\bw^*,\bq) = \varphi/2$, giving
\begin{equation}
  h(\bw^*) \;=\;
    \frac{\|\bp\|_{\SZ}\,\|\bq\|_{\SZ}}{\|\bA\|^2}
    \;\cdot\; \frac{1+\cos\varphi}{2}.
  \label{eq:hstar}
\end{equation}
\end{proposition}

The proof projects $h(\bw)$ onto the two-dimensional span of $\bp$ and
$\bq$, reduces the problem to a one-dimensional trigonometric optimisation,
and solves it in closed form (see the Supplementary Material).
The bisector result has a clean geometric interpretation: $h(\bw)$ rewards a weight vector simultaneously well-aligned
with both path directions $\bp$ (the treatment path) and $\bq$ (the outcome
path).  Leaning all the way toward $\bp$ maximises alignment with the
treatment path but sacrifices alignment with the outcome path, and vice
versa; the bisector $\bw^*$ achieves the best symmetric compromise.
Formally, the direction that maximises $\hat\alpha(\bw)$ alone is
$\bw\propto\bp$, and the direction that maximises $\hat\beta(\bw)$ alone
is $\bw\propto\bq$; at either single-path extreme the alignment is
$\cos\varphi$, whereas the bisector achieves $(1+\cos\varphi)/2 >
\cos\varphi$ for any $\varphi>0$.  The gain $(1+\cos\varphi)/2 -
\cos\varphi = (1-\cos\varphi)/2$ is largest when the two paths are nearly
orthogonal ($\varphi\approx\pi/2$), precisely the regime where single-path
methods fail most severely.  This is not merely a computational shortcut:
the bisector result establishes that jointly optimising both mediation paths
is always strictly better than optimising either alone, with the improvement
increasing as the paths diverge.

\subsection{Back-Transformation and Implementation}

The optimal weight vector is
\begin{equation}
  \bw^* \;\propto\;
    \frac{\bp}{\|\bp\|_{\SZ}}
    \;+\;
    \frac{\bq}{\|\bq\|_{\SZ}},
  \label{eq:reconstruct_ab}
\end{equation}
the sum of the two unit path directions in the $\SZ$-metric.
Since $\bw^*$ depends only on the directions of $\bp$ and $\bq$, it is invariant to rescaling of $\bA$ or $\bY$.
To obtain weights that are directly comparable across features with different units or variances---for example when ranking proteins by their contribution to the composite mediator---it suffices to column-standardise $\bX$ before running the solver; the resulting weights are equivalent to $w^*_j \cdot \hat\sigma_j$ and measure the change in $M$ per one-standard-deviation increase in each feature.
Computing $\bp$ and $\bq$ requires factorising $\SZ + \varepsilon I$ once
($\calO(p^3/3)$) and two back-substitutions ($\calO(p^2)$ each), for a total
cost of $\calO(np + p^3/3)$; $\SZ^{-1/2}$ is never formed.
Because $\theta^* = \varphi/2$ is available in closed form, no grid search
is needed (contrast MaxCor, which optimises the mediation index $f^*(\bw)=\cor(\bX\bw,\bA)\cdot\cor(\bZ\bw,\bY)$ via a semi-closed-form solution; see the Supplementary Material).

The suppression mediator $\bw^*_{-}$~\eqref{eq:objective_supp} is obtained
by replacing $\bz$ with $-\bz$, which negates $\bq = \SZ^{-1}\bz$ at no
extra cost; both $\bw^*_{+}$ and $\bw^*_{-}$ are returned by a single call.
After orienting each solution so that $\hat\alpha(\bw^*_{\pm}) \geq 0$,
the sign of $\hat\beta$ identifies the effect type:
$\hat\beta > 0$ for concordant mediation, $\hat\beta < 0$ for suppression.

\begin{proposition}[Consistency of $\hat\bw^*_+$]
\label{prop:consistency}
Under the conditions of Proposition~\ref{prop:power} with both paths active,
where $\balpha_0 = n^{-1}E[\bX^\top\bA]$ and $\bbeta_0 = n^{-1}E[\bZ^\top\bY]$
are the population path vectors
($\balpha_0 \neq \mathbf{0}$, $\bbeta_0 \neq \mathbf{0}$),
the sample maximiser $\hat\bw^*_+$ satisfies
$\hat\bw^*_+ / \|\hat\bw^*_+\|_{\SZ} \xrightarrow{p}
\bw^*_{0} / \|\bw^*_{0}\|_{\Sigma_Z}$ as $n\to\infty$ with $p$ fixed,
where $\Sigma_Z = n^{-1}E[\SZ]$ is the population Gram matrix and
$\bw^*_{0}$ bisects the population path vectors
$\Sigma_Z^{-1}\balpha_0$ and $\Sigma_Z^{-1}\bbeta_0$ in the $\Sigma_Z$-metric.
\end{proposition}

The proof uses convergence of the sufficient statistics and continuity of the bisector map (see the Supplementary Material).
Intuitively, MaxIE depends on the data only through $n^{-1}\ba$, $n^{-1}\bz$, and $n^{-1}\SZ$, each converging to its population counterpart; smoothness of the bisector map then gives consistency whenever both path vectors are nonzero.

\section{A Global Test for the Existence of Any Composite Mediator}
\label{sec:inference}

A natural question is whether any composite mediator exists in the population.
Letting $\alpha(\bw)$ and $\beta(\bw)$ denote the population counterparts of
$\hat\alpha(\bw)$ and $\hat\beta(\bw)$~\eqref{eq:pathcoeffs}, defined by
\begin{equation}
  \alpha(\bw) \;=\; \frac{E[\bA^\top(\bX\bw)]}{E[\|\bA\|^2]}, \qquad
  \beta(\bw) \;=\; \frac{E[(\bZ\bw)^\top\bY]}{E[\|\bZ\bw\|^2]},
  \label{eq:pop_paths}
\end{equation}
the global null is
\begin{equation}
  H_0:\; \max_{\bw}\,|\alpha(\bw)\,\beta(\bw)| \;=\; 0,
  \label{eq:H0}
\end{equation}
i.e.\ no linear combination of $\bX$ produces a non-zero indirect effect ---
concordant \emph{or} suppression --- in the population.
By~\eqref{eq:pop_paths}, this is equivalent to
\begin{equation}
  E[\bX^\top\bA] = \mathbf{0} \quad\text{or}\quad E[\bZ^\top\bY] = \mathbf{0}.
  \label{eq:H0_moments}
\end{equation}
A natural candidate test statistic is $h(\bw^*)$, the maximised sample indirect
effect.  However, $h(\bw^*)$ is always strictly positive in finite samples and
its distribution under $H_0$ depends on the magnitude of the active path ---
an unknown nuisance --- so it cannot be pivoted into a valid test.

The factorisation~\eqref{eq:hstar} shows that
\begin{equation*}
  h(\bw^*) \;=\;
    \frac{\|\bp\|_{\SZ}\,\|\bq\|_{\SZ}}{\|\bA\|^2}
    \cdot \frac{1 + \cos\varphi}{2},
\end{equation*}
where $\cos\varphi$ is the cosine of the angle between the two path vectors
in the $\SZ$-metric,
\begin{equation}
  \cos\varphi \;=\; \frac{\bp^\top\SZ\,\bq}{\|\bp\|_\SZ\,\|\bq\|_\SZ}
    \;=\; \frac{\ba^\top\bq}{\sqrt{\ba^\top\bp}\,\sqrt{\bz^\top\bq}},
  \label{eq:costest}
\end{equation}
where the last equality uses $\SZ\bp=\ba$ and $\SZ\bq=\bz$, so the statistic
is a free by-product of the MaxIE solver requiring no additional computation.
The first factor of $h(\bw^*)$ is the path-strength nuisance;
$\cos\varphi$ depends only on the directions of the two paths, not their
magnitudes.
Under $H_0$, one path direction is uniformly distributed on $S^{p-1}$
(shown in the proof below), making $\cos\varphi$ a pivot.

\begin{proposition}[Null distribution of $\cos\varphi$]
\label{prop:costest}
Suppose Assumptions A1--A4 hold, $\bA \mid \bX \sim
\mathcal{N}(E[\bA\mid\bX], \sigma_A^2 I_n)$, and
$\bY \mid \bX, \bA \sim \mathcal{N}(E[\bY\mid\bX,\bA], \sigma_Y^2 I_n)$
for unknown $\sigma_A^2, \sigma_Y^2 > 0$.
Results are stated in the asymptotic regime $n \to \infty$ with $p$ fixed.
Then under $H_0$,
\begin{equation}
  \cos^2\!\varphi \;\sim\; \mathrm{Beta}\!\left(\tfrac{1}{2},\,\tfrac{p-1}{2}\right)
  \label{eq:null_dist}
\end{equation}
asymptotically as $n\to\infty$ with $p$ fixed.
Equivalently, the signed statistic
\begin{equation}
  T \;=\; \cos\varphi \cdot \sqrt{\frac{p-1}{1-\cos^2\!\varphi}}
  \;\sim\; t(p-1),
  \label{eq:Tstat}
\end{equation}
where the sign of $T$ matches that of $\cos\varphi$.
The test rejects $H_0$ when $|T| > t_{1-\alpha/2,\,p-1}$ and has size $\alpha$
asymptotically.  Two-sided and directional one-sided tests (for concordant and
suppression mediation) follow immediately from the sign of $T$, all at no
extra computational cost.
\end{proposition}

The proof establishes that whichever path is null under $H_0$, its
whitened direction is uniform on $S^{p-1}$: exactly so when the $\alpha$-path is null
($E[\bX^\top\bA]=\mathbf{0}$), and asymptotically
as $n\to\infty$ with $p$ fixed when the $\beta$-path is null
($E[\bZ^\top\bY]=\mathbf{0}$).  The Beta distribution of
$\cos^2\!\varphi$ then follows from the standard result on the inner
product of a fixed and a uniform unit vector; see the Supplementary Material.

Notably, the two null scenarios are not symmetric: the $t(p-1)$
distribution is \emph{exact} (for any $n>p$) when the $\beta$-path is null,
but only \emph{asymptotic} in $n$ when the $\alpha$-path is null.
The same asymmetry holds in the dual (Proposition~\ref{prop:dual}):
exact for any $p>n$ when the $\beta$-path is null, asymptotic as $p\to\infty$
when the $\alpha$-path is null.
This is visible in the finite-sample size results of Table~\ref{tab:inference}.

The assumption $n > p$ is required for the primal test: when $p \geq n$,
$\SZ$ is rank-deficient and the path directions are no longer uniform on
$S^{p-1}$.  The dual formulation of Proposition~\ref{prop:dual} extends
the test to $p \geq n$ with degrees of freedom $n-2$ in place of $p-1$.

The intuition behind the result is transparent once the null is stated
geometrically.  Under $H_0$, at least one of the two path vectors carries
no population signal, so its sample direction is essentially noise --- a
random unit vector in $\mathbb{R}^p$ drawn independently of the other path.
The cosine of the angle between a fixed unit vector and an independent
random unit vector has a well-known distribution on $[-1,1]$, symmetric
about zero and concentrated near zero for large $p$.
The $t(p-1)$ statistic is simply the natural pivot for this cosine, with
the degrees of freedom $p-1$ reflecting the dimensionality of the sphere.
Crucially, $\cos\varphi$ is already computed by MaxIE during the bisector
step --- the test requires no additional passes through the data.

\begin{proposition}[Power of the cosine test]
\label{prop:power}
Under Assumptions A1--A4 and the conditions of Proposition~\ref{prop:costest}, let
$\Sigma_Z = n^{-1}E[\SZ]$, and define the population angle
$\varphi_0\in(0,\pi)$ between the two path directions in the
$\Sigma_Z^{-1}$-metric,
\begin{equation}
  \cos\varphi_0 \;=\; \frac{\balpha_0^\top\Sigma_Z^{-1}\bbeta_0}
    {\sqrt{\balpha_0^\top\Sigma_Z^{-1}\balpha_0\;}\,
     \sqrt{\bbeta_0^\top\Sigma_Z^{-1}\bbeta_0\;}}.
  \label{eq:phi0}
\end{equation}
As $n\to\infty$ with $p$ fixed, $\cos\varphi\xrightarrow{p}\cos\varphi_0$
and $T\xrightarrow{p}\delta$ where
\begin{equation}
  \delta(\varphi_0,p) \;=\; \cot\varphi_0\cdot\sqrt{p-1}.
  \label{eq:nct}
\end{equation}
The asymptotic power of the two-sided test satisfies:
(i)~the test has size $\alpha$ under $H_0$;
(ii)~power is a U-shaped function of $\varphi_0$, strictly
decreasing on $(0,\pi/2)$ and strictly increasing on $(\pi/2,\pi)$,
with equal power at $\varphi_0$ and $\pi-\varphi_0$;
(iii)~power tends to $1$ as $n\to\infty$ if and only if
$|\delta(\varphi_0,p)|>t_{1-\alpha/2,\,p-1}$.
\end{proposition}

The proof applies convergence of the sufficient statistics and continuity of $T$
to show $T\xrightarrow{p}\delta$, then reads off each property from the
monotonicity of $|\cot\varphi_0|$ (see the Supplementary Material).
The noncentrality $\delta = \cot\varphi_0\cdot\sqrt{p-1}$ ties power
directly to the geometry of the path directions: it is the cotangent of
the population angle scaled by $\sqrt{p-1}$, large when $\varphi_0$ is
close to $0$ or $\pi$ and zero at $\varphi_0=\pi/2$.  The power curve
is therefore U-shaped: the test is most powerful when both path directions
nearly coincide ($\varphi_0\approx 0$, concordant mediation) or nearly
oppose ($\varphi_0\approx\pi$, suppression), and powerless when they are
orthogonal.  The two configurations are equally detectable at the same
angular distance from $\pi/2$, since $|\cot(\pi-\varphi_0)|=|\cot\varphi_0|$.

The detection threshold is $|\delta(\varphi_0,p)|=t_{1-\alpha/2,p-1}$,
equivalently $|\cot\varphi_0|=t_{1-\alpha/2,p-1}/\sqrt{p-1}$.  For
fixed $p$, if $|\delta|$ falls below this threshold, increasing $n$
cannot help once $T$ has concentrated at $\delta$ --- a qualitative
difference from standard tests where power always increases
with $n$.

Non-rejection of $H_0$ therefore has three distinct causes:
(1)~no mediation ($\balpha_0=\mathbf{0}$ or $\bbeta_0=\mathbf{0}$);
(2)~paths present but $|\delta(\varphi_0,p)|\leq t_{1-\alpha/2,p-1}$,
where no increase in $n$ will help; or
(3)~$n$ not yet large enough for $T$ to concentrate near $\delta$,
where collecting more data helps until the asymptotic regime is reached.
Empirical validation is given in Section~\ref{sec:inf_sim}.

\section{Dual Implementation of MaxIE for $p \geq n$}
\label{sec:dual}

When $p \geq n$, the primal Gram matrix $\SZ = \bZ^\top\bZ\in\mathbb{R}^{p\times p}$
has rank $n-1 < p$ and is therefore singular: the primal path vectors
$\bp = \SZ^{-1}\ba$ and $\bq = \SZ^{-1}\bz$ are undefined, and the
primal cosine test breaks down.
The dual implementation resolves this by working in $\mathbb{R}^n$ instead of
$\mathbb{R}^p$, replacing $\SZ$ with the $n\times n$ kernel matrix
$\bK_Z = \bZ\bZ^\top$ and projecting the path vectors into the column space
of $\bZ$.

The foundation is the identity $\bK_Z^+\,\bZ = \bZ\,\SZ^+$, which gives
dual path vectors
\begin{equation}
  \tilde\bp \;:=\; \bZ\bp \;=\; \bK_Z^+\tilde\ba, \qquad
  \tilde\bq \;:=\; \bZ\bq \;=\; \bK_Z^+\tilde\bz,
  \label{eq:dual_pq}
\end{equation}
where $\tilde\ba = \bK\bA$ and $\tilde\bz = \bK\bQ_A\bY$ are dual sufficient
statistics computable without forming $\SZ$,
\begin{equation}
  \tilde\ba \;=\; \bK\bA, \qquad \tilde\bz \;=\; \bK\bQ_A\bY.
  \label{eq:dual_stats}
\end{equation}
Since $\tilde\bp=\bZ\bp$ and $\tilde\bq=\bZ\bq$, the Euclidean angle
between them in $\mathbb{R}^n$ equals the primal $\SZ$-metric angle:
$\cos\varphi = \tilde\bp^\top\tilde\bq/(\|\tilde\bp\|\|\tilde\bq\|)$.
The ambient dimension is $n-1$ (the rank of $\bK_Z$) rather than $p-1$,
so $T\sim t(n-2)$ under $H_0$ instead of $t(p-1)$.
The primal weight vector is recovered as
\begin{equation}
  \bw^* \;\propto\;
    \bZ^+\!\left(
      \frac{\tilde\bp}{\|\tilde\bp\|}+\frac{\tilde\bq}{\|\tilde\bq\|}
    \right),
  \label{eq:dual_recovery}
\end{equation}
where $\bZ^+ = \bZ^\top\bK_Z^+$, at cost $O(np)$.

\begin{proposition}[Dual implementation of MaxIE]
\label{prop:dual}
Suppose $\bX$ has full row rank (satisfied when $p \geq n$ and $\bX$ is in
general position).  Let $(\tilde\ba,\tilde\bz)$ be as in~\eqref{eq:dual_stats}
and $(\tilde\bp,\tilde\bq)$ as in~\eqref{eq:dual_pq}.
\begin{enumerate}[label=(\roman*)]
  \item \emph{(Angle preservation and null distribution.)}
    The dual path vectors satisfy $\tilde\bp=\bZ\bp$ and $\tilde\bq=\bZ\bq$,
    so their Euclidean angle equals the primal angle:
    \begin{equation}
      \cos\varphi \;=\;
        \frac{\tilde\bp^\top\tilde\bq}{\|\tilde\bp\|\,\|\tilde\bq\|}
        \;=\;
        \frac{\bp^\top\SZ\bq}{\|\bp\|_{\SZ}\|\bq\|_{\SZ}}.
      \label{eq:dual_costest}
    \end{equation}
    Under the Gaussian assumptions of Proposition~\ref{prop:costest} and $H_0$,
    \begin{equation}
      T \;=\; \cos\varphi\cdot
        \sqrt{\frac{n-2}{1-\cos^2\!\varphi}}
        \;\sim\; t(n-2)
      \label{eq:dual_Tstat}
    \end{equation}
    asymptotically as $p\to\infty$ (with $n$ fixed) under $H_0$.
    The test rejects $H_0$ at size $\alpha$ when $|T|>t_{1-\alpha/2,\,n-2}$.  \item \emph{(Consistency of the dual path vectors.)}
    Suppose $\bX$ has i.i.d.\ columns with $E[\bX_j]=\mathbf{0}$ and
    $E[\bX_j\bX_j^\top]=\Sigma_X\succ 0$.
    Define the population dual sufficient statistics
    $\tilde\balpha_0 = \Sigma_X\bA$ and $\tilde\bbeta_0 = \Sigma_X\bQ_A\bY$,
    the population kernel $\Sigma_K = \bQ_A\Sigma_X\bQ_A$ (rank $n-1$),
    and the population dual path vectors
    \begin{equation}
      \tilde\bp_0 \;=\; \Sigma_K^+\tilde\balpha_0, \qquad
      \tilde\bq_0 \;=\; \Sigma_K^+\tilde\bbeta_0,
      \label{eq:dual_pop_pq}
    \end{equation}
    both in $\bA^\perp\subset\mathbb{R}^n$.
    If both are nonzero, then as $p\to\infty$ with $n$ and $(\bA,\bY)$ fixed,
    \[
      \frac{\tilde\bp}{\|\tilde\bp\|} \;\xrightarrow{p}\;
      \frac{\tilde\bp_0}{\|\tilde\bp_0\|}, \qquad
      \frac{\tilde\bq}{\|\tilde\bq\|} \;\xrightarrow{p}\;
      \frac{\tilde\bq_0}{\|\tilde\bq_0\|},
    \]
    and consequently $\cos\varphi\xrightarrow{p}\cos\phi_0(\bA,\bY)$
    as defined in item~(iii) below.
  \item \emph{(Power of the dual cosine test.)}
    Assume $\tilde\bp_0\neq\mathbf{0}$ and $\tilde\bq_0\neq\mathbf{0}$
    (both mediation paths active at the population level).
    Let $\phi_0(\bA,\bY)=\arccos\!\left(\tilde\bp_0^\top\tilde\bq_0/(\|\tilde\bp_0\|\|\tilde\bq_0\|)\right)\in(0,\pi)$
    denote the population angle between the dual path vectors;
    as $p\to\infty$ with $n$ and $(\bA,\bY)$ fixed,
    \begin{equation}
      \cos\varphi \;\xrightarrow{p}\; \cos\phi_0(\bA,\bY),
      \label{eq:cosphi_limit}
    \end{equation}
    a random variable since $(\bA,\bY)$ remain random with $n$ fixed.
    The statistic $T$ under $H_1$ is approximated by a noncentral $t$ with $n-2$ degrees of
    freedom and noncentrality
    \begin{equation}
      \delta(\bA,\bY) \;=\; \cot\!\phi_0(\bA,\bY)\cdot\sqrt{n-2},
      \label{eq:dual_noncentrality}
    \end{equation}
    with limiting power
    \begin{equation}
      \pi_\infty \;\approx\;
        P\!\left(\left|t\!\left(n-2,\,\delta(\bA,\bY)\right)\right|
          \geq t_{1-\alpha/2,\,n-2}\;\Big|\;\bA,\bY\right).
      \label{eq:pi_inf}
    \end{equation}
    Consequently: (a)~$T\approx t(n-2,\delta(\bA,\bY))$ under $H_1$,
    confirmed numerically (Figure~\ref{fig:power_dual}, left);
    (b)~power is U-shaped in $\phi_0$, minimised at $\pi/2$ and
    symmetric about it (Figure~\ref{fig:power_dual}, centre);
    (c)~power converges to $\pi_\infty\in(\alpha,1)$ for all
    $\phi_0\neq\pi/2$, with no sharp detection threshold in $p$
    (Figure~\ref{fig:power_dual}, right).
\end{enumerate}
The dual algorithm costs $\calO(n^2 p + n^3/3)$: $\calO(np)$ to form
$\bZ = \bQ_A\bX$, $\calO(n^2 p)$ to form $\bK_Z = \bZ\bZ^\top$,
and $\calO(n^3/3)$ for its eigendecomposition plus $\calO(n^2)$ for
two back-substitutions to obtain $(\tilde\bp,\tilde\bq)$, and $\calO(np)$
for weight recovery~\eqref{eq:dual_recovery}.
\end{proposition}

Item~(i) establishes angle preservation and the $t(n-2)$ null distribution.
The null distribution is \emph{asymptotic} in both regimes but in different directions:
in the primal ($n\to\infty$, $p$ fixed) the $t(p-1)$ approximation improves as path
directions stabilise; in the dual ($p\to\infty$, $n$ fixed) the $t(n-2)$ approximation
improves as the dual path vectors stabilise, with finite-$p$ distortions from
each null scenario vanishing as $p\to\infty$.
Item~(ii) establishes consistency of $\cos\varphi$ as an estimator of the
population angle; item~(iii) characterises power via a noncentral $t(n-2)$,
with the key distinction from the primal that power converges to a finite
limit $\pi_\infty\in(\alpha,1)$ rather than 1, because $\cos\varphi$ stabilises
at the random variable $\cos\phi_0(\bA,\bY)$ rather than a fixed constant.

The primal and dual are two asymptotic views of the same $\cos\varphi$:
primal fixes $p$ and lets $n\to\infty$ (point-mass limit),
dual fixes $n$ and lets $p\to\infty$ (non-degenerate limit).
At the boundary $p=n$, the noncentralities $\cot\varphi_0\sqrt{p-1}$ and
$\cot\phi_0\sqrt{n-2}$ differ by $\sqrt{(n-2)/(n-1)}\to 1$, a negligible effect.

Table~S1 of the Supplementary Material summarises both implementations:
they are not two methods but one method in two coordinate systems.
The natural choice is whichever is feasible: primal ($\calO(p^3/3)$, $O(p^2)$ memory)
when $p<n$, dual ($\calO(n^2 p+n^3/3)$, $O(n^2)$ memory) when $p\geq n$.

\section{Comparison Methods}
\label{sec:baselines}

We compare MaxIE against two OLS-based methods and the Directions of
Mediation (DM) method of Chén et al.~\citep{chen2018dm}, and include MaxCor as an
additional comparison to examine how optimising $f^*$ rather than
$\hat\alpha\hat\beta$ affects performance. Method numbers match those used in
the experiment tables.

\textbf{[1] MaxIE} is our primary proposed method, directly optimising
$\hat\alpha\hat\beta$ via the closed-form solution of Section~\ref{sec:indirect}.
\textbf{[2] MaxCor} is a comparison solver that directly optimises the
\emph{mediation index} $f^*(\bw) = \cor(\bX\bw,\bA)\cdot\cor(\bZ\bw,\bY)$,
a bounded, scale-free alternative to $\hat\alpha\hat\beta$ that remains
well-defined when the total effect $\hat\tau\approx 0$
(the Supplementary Material); the semi-closed-form solution is in
the Supplementary Material.

\textbf{[3] Reg Y\textasciitilde X} sets $\bw = \SX^{-1}\bX^\top\bY
= \SX^{-1}(\bz + \frac{\bA^\top\bY}{\|\bA\|^2}\ba)$, where
$\SX = \bX^\top\bX$ is the full Gram matrix, so this direction
maximises $\cor(\bX\bw,\bY)$ without controlling for $\bA$. This corresponds
to the standard outcome-prediction model.

\textbf{[4] Reg A\textasciitilde X} sets $\bw = \SX^{-1}\ba$, the direction
that maximises $\cor(\bX\bw,\bA)$ alone, ignoring the outcome $\bY$. This
corresponds to the standard approach for constructing biological age clocks in
aging research, where a composite molecular score is trained to predict
chronological age without reference to any disease outcome.

\textbf{[5] DM} (Directions of Mediation;~\citep{chen2018dm}) finds the loading
vector $\bw$ by maximising the LSEM likelihood via an iterative eigenvalue
algorithm.  The first direction of mediation (DM1) is the composite $\bX\bw$
that accounts for the largest share of the joint log-likelihood of the two
path models.  Unlike MaxIE, DM has no closed-form solution and requires
iterative optimisation; the two methods share the same conceptual objective
--- finding a single composite mediator that best captures the indirect effect
--- but differ in criterion (likelihood vs.\ $\hat\alpha\hat\beta$) and
algorithm (iterative vs.\ closed-form).  We implemented DM via a native
reimplementation of the \texttt{hdmed} algorithm~\citep{clark2023,chen2018dm}.

Methods [3] and [4] each optimise a single path coefficient, leaving the
other path unconstrained. In addition to comparing mediation performance
across all methods, we report wall-clock runtimes for MaxIE and MaxCor
against their numerical solver counterparts (L-BFGS-B with multiple random
restarts applied to $\hat\alpha\hat\beta$ and $f^*$ respectively) to quantify
the practical benefit of the closed-form solutions.

For the inference simulations of Section~\ref{sec:inf_sim}, we compare the
two-sided cosine test against the \textbf{intersection-union test (IUT)}.
The IUT rejects $H_0$ only when both the omnibus $F$-test for
$R^2(\bA \sim \bX)=0$ and the incremental $F$-test for
$R^2(\bY \sim \bZ)=0$ reject individually;
its $p$-value is $\max(p_\alpha, p_\beta)$.  Size is controlled over $H_0$
by the intersection-union principle~\citep{berger1982}.

\section{Experiments}
\label{sec:experiments}

\subsection{Optimisation}

We compare MaxIE and MaxCor against the OLS baselines [3] and [4] across
four scenarios that vary the alignment between $\balpha$ and $\bbeta$.
We simulate $n$ observations from
\begin{align}
  \bX &\;\sim\; \mathcal{N}(0,\Sigma_{\mathrm{AR1}}), \notag\\
  \bA &\;=\; \bX\balpha + \beps_A, \qquad \beps_A \sim \mathcal{N}(0,0.25\,I_n),
  \label{eq:dgp}\\
  \bY &\;=\; \bX\bbeta + \tau\bA + \beps_Y, \qquad \beps_Y \sim \mathcal{N}(0,0.25\,I_n),
  \notag
\end{align}
where $\balpha, \bbeta \in \mathbb{R}^p$ are the true population path vectors,
$\tau = 0.25$, $[\Sigma_{\mathrm{AR1}}]_{ij} = \rho^{|i-j|}$ with
$\rho = 0.75$, and rows of $\bX$ are drawn i.i.d.\ and column-centred.
Results are averaged over 20 replicates; $\SZ$ is computed
from~\eqref{eq:sigmas} without forming $\bZ$ explicitly.

In all four scenarios $\balpha$ and $\bbeta$ are sparse unit vectors
($\|\balpha\|=\|\bbeta\|=1$) with $p/4$ nonzero entries drawn from
$\mathcal{N}(0,1)$, zero elsewhere; the scenarios differ only in how much
their supports overlap.
Concretely, draw a random permutation of $\{1,\ldots,p\}$ and partition it
into a \emph{shared} index set of size $s$ and two \emph{unique} sets of
size $p/4 - s$ each; assign i.i.d.\ $\mathcal{N}(0,1)$ values on the shared
indices to both vectors identically, and independent $\mathcal{N}(0,1)$
values on the unique indices, then normalise each to unit length.
The four scenarios set $s\in\{0, p/16, p/8, p/4\}$, labelled
\textbf{S1} ($s=0$, independent, $\cos\approx 0$),
\textbf{S2} ($s=p/16$, 25\% overlap, $\cos\approx 0.23$),
\textbf{S3} ($s=p/8$, 50\% overlap, $\cos\approx 0.47$), and
\textbf{S4} ($s=p/4$, $\bbeta=\balpha$, identical, $\cos=1$).
All shared entries are drawn from the same random vector, so
$\balpha^\top\bbeta \geq 0$ with high probability; the scenarios therefore
cover concordant mediation only.
The suppression solver ($\bw^*_-$) is validated by symmetry: negating $\bY$
maps every concordant instance to an equivalent suppression instance.
Tables~\ref{tab:obj_h} and~S2 report mean $h = \hat\alpha(\bw^*)\hat\beta(\bw^*)$
and $f^* = f^*(\bw^*)$ respectively across the four scenarios and $(n,p)$ settings, with
standard deviations over 20 replicates.

\begin{table}[htbp]
\centering
\caption{Mean indirect effect $h = \hat\alpha\hat\beta$ with standard deviation
in parentheses, averaged over 20 replicates. Bold indicates the best value per row.
S1: $\balpha\perp\bbeta$ (independent, $\cos\approx0$);
S2: 25\% support overlap ($\cos\approx0.23$);
S3: 50\% support overlap ($\cos\approx0.47$);
S4: $\balpha=\bbeta$ ($\cos=1$).
[5]~DM: native reimplementation of \texttt{hdmed::mediate\_hdmm}~\citep{clark2023,chen2018dm}.}
\label{tab:obj_h}
\smallskip
\resizebox{\linewidth}{!}{%
\setlength{\tabcolsep}{6pt}
\begin{tabular}{llr rrrr r}
\toprule
Scenario & $n$ & $p$
  & \multicolumn{1}{c}{[1] MaxIE}
  & \multicolumn{1}{c}{[2] MaxCor}
  & \multicolumn{1}{c}{[3] Reg Y\textasciitilde X}
  & \multicolumn{1}{c}{[4] Reg A\textasciitilde X}
  & \multicolumn{1}{c}{[5] DM} \\
\midrule
S1: $\balpha\perp\bbeta$
  &  500 &   20 & \textbf{0.86}\,(0.19) & 0.80\,(0.17) & $0.23$\,(0.28) & $0.04$\,(0.28) & $0.12$\,(0.72) \\
  & 1000 &  100 & \textbf{0.92}\,(0.14) & 0.84\,(0.12) & $0.18$\,(0.13) & $-0.02$\,(0.14) & $-0.04$\,(0.74) \\
  & 1000 &  500 & \textbf{1.41}\,(0.09) & 1.21\,(0.08) & $0.21$\,(0.06) & $-0.01$\,(0.07) & $-0.44$\,(0.75) \\
  &   500 &  1000 & \textbf{1.02}\,(0.06) & \multicolumn{4}{c}{\textit{n/a \textemdash{} dual implementation ($p \geq n$)}} \\ 
\midrule
S2: 25\% overlap
  &  500 &   20 & \textbf{0.92}\,(0.21) & 0.85\,(0.19) & $0.41$\,(0.23) & $0.23$\,(0.25) & $0.47$\,(0.63) \\
  & 1000 &  100 & \textbf{0.99}\,(0.13) & 0.91\,(0.12) & $0.38$\,(0.13) & $0.19$\,(0.13) & $0.63$\,(0.32) \\
  & 1000 &  500 & \textbf{1.46}\,(0.08) & 1.24\,(0.06) & $0.38$\,(0.09) & $0.16$\,(0.10) & $0.75$\,(0.56) \\
  &   500 &  1000 & \textbf{1.04}\,(0.09) & \multicolumn{4}{c}{\textit{n/a \textemdash{} dual implementation ($p \geq n$)}} \\ 
\midrule
S3: 50\% overlap
  &  500 &   20 & \textbf{0.89}\,(0.17) & 0.82\,(0.15) & $0.59$\,(0.16) & $0.43$\,(0.18) & $0.80$\,(0.23) \\
  & 1000 &  100 & \textbf{1.02}\,(0.09) & 0.94\,(0.07) & $0.54$\,(0.13) & $0.37$\,(0.14) & $0.87$\,(0.13) \\
  & 1000 &  500 & \textbf{1.45}\,(0.08) & 1.24\,(0.06) & $0.58$\,(0.06) & $0.36$\,(0.09) & $1.07$\,(0.12) \\
  &   500 &  1000 & \textbf{1.01}\,(0.07) & \multicolumn{4}{c}{\textit{n/a \textemdash{} dual implementation ($p \geq n$)}} \\ 
\midrule
S4: $\balpha=\bbeta$
  &  500 &   20 & \textbf{0.80}\,(0.06) & 0.79\,(0.06) & 0.79\,(0.06) & 0.78\,(0.06) & \textbf{0.80}\,(0.06) \\
  & 1000 &  100 & \textbf{0.87}\,(0.05) & 0.85\,(0.05) & 0.84\,(0.05) & 0.82\,(0.05) & \textbf{0.87}\,(0.05) \\
  & 1000 &  500 & \textbf{1.16}\,(0.05) & 1.02\,(0.04) & $0.93$\,(0.03) & 0.80\,(0.06) & $1.12$\,(0.04) \\
  &   500 &  1000 & \textbf{0.81}\,(0.03) & \multicolumn{4}{c}{\textit{n/a \textemdash{} dual implementation ($p \geq n$)}} \\ 
\bottomrule
\end{tabular}%
}
\end{table}

\paragraph{Optimisation results.}
Tables~\ref{tab:obj_h} and~S2 confirm that the two proposed
solvers are complementary. MaxCor dominates on $f^*$ across all scenarios
and settings; MaxIE dominates on $h = \hat\alpha\hat\beta$ by a consistent
margin of $0.05$--$0.21$, most pronounced at large $p$.
The two objectives are genuinely distinct, consistent with the theory in the
Supplementary Material.
MaxCor does not admit a dual implementation for $p \geq n$ because $\SX$
is rank-deficient in this regime; the corresponding entries are marked n/a.
The single-arm baselines [3] and [4] perform well only in S4 ($\balpha=\bbeta$),
where optimising one path implicitly optimises both; they degrade progressively
as path alignment decreases, reaching near-zero or negative $h$ in S1.
DM matches MaxIE closely in S4 but degrades as overlap decreases, performing
poorly in S1 ($h$ near zero, std $\approx0.7$--$0.8$) where conflicting
path gradients cause the iterative solver to converge unpredictably.
MaxIE is therefore not only faster but strictly better on $h$ across all
settings, with the largest advantage when $\balpha$ and $\bbeta$ are
poorly aligned.

Both closed-form solvers share the same $\mathcal{O}(np + p^3/3)$ cost
structure: $\mathcal{O}(np)$ for sufficient statistics, $\mathcal{O}(p^3/3)$
for a Cholesky factorisation, and $\mathcal{O}(p^2)$ for two triangular
solves. This is the same asymptotic cost as the OLS methods [3] and [4],
which each require a single linear solve of the same order. In practice,
MaxIE and MaxCor run at the same order of magnitude as [3] and [4],
meaning that jointly optimising the mediation objective adds essentially no
computational overhead compared to the single-path OLS regressions.
Table~\ref{tab:time} pairs each closed-form solver against its numerical
counterpart (L-BFGS-B, 10 restarts) at $p\in\{20,100\}$.  The objective panel
shows that [1] MaxIE and Num-$h$ achieve identical values of $h$, and [2] MaxCor
and Num-$f^*$ achieve identical values of $f^*$, to within simulation noise
across all replications.  This confirms that the closed-form solutions are
globally optimal: the analytic derivation loses nothing in objective quality
relative to an unconstrained numerical search.  The timing panel covers all
three $(n,p)$ configurations.
The closed-form speedup over L-BFGS-B is approximately $400\times$ at $p=20$
($0.4$\,ms vs.\ $143$\,ms for the $h$ pair; $0.6$\,ms vs.\ $489$\,ms for $f^*$)
and grows to roughly $200\times$ at $p=100$
($9$\,ms vs.\ $1{,}952$\,ms for $h$; $9$\,ms vs.\ $19{,}278$\,ms for $f^*$).
The regression baselines [3] and [4] match MaxIE and MaxCor in speed across all
settings, as all four reduce to a single Cholesky solve
($0.3$\,ms at $p=20$ up to $218$\,ms at $p=500$).
Among the baselines, DM is the slowest: $7.5$\,ms at $p=20$, $56$\,ms at
$p=100$, and $580$\,ms at $p=500$, reflecting its iterative eigendecomposition.
All experiments were run on a single CPU core (\texttt{Apple M1 Pro},
\texttt{32\,GB} RAM).

\begin{table}[htbp]
\centering
\caption{Closed-form solvers vs.\ numerical baselines (L-BFGS-B, 10 restarts),
DM (iterative LSEM), and regression baselines.
Same DGP as Tables~\ref{tab:obj_h}--S2 (AR(1), $\rho=0.75$, $\tau=0.25$,
$\sigma_\epsilon=0.5$), scenario S3 (50\% overlap), averaged over 5 replicates.
\textit{Top panel}: mean objective value (sd) at $(n,p)\in\{(500,20),(1000,100)\}$;
[1]~MaxIE matches Num-$h$ and [2]~MaxCor matches Num-$f^*$ to within simulation
noise, confirming the closed-form solutions are globally optimal.
Dashes (---) indicate baselines that do not directly optimise $h$ or $f^*$.
\textit{Bottom panel}: mean wall-clock time (ms) at
$(n,p)\in\{(500,20),(1000,100),(1000,500),(500,1000)\}$;
\textit{not run} indicates numerical methods omitted at $p=500$ due to prohibitive runtime;
the $p=1000$ row uses the dual implementation.
All results obtained on a single CPU core (\texttt{Apple M1 Pro}, \texttt{32\,GB} RAM).}
\label{tab:time}
\smallskip
\resizebox{\linewidth}{!}{%
\setlength{\tabcolsep}{5pt}
\begin{tabular}{rr cc cc c ccc}
\toprule
 & & \multicolumn{2}{c}{Optimise $h = \hat\alpha\hat\beta$}
   & \multicolumn{2}{c}{Optimise $f^*$}
   & & \multicolumn{3}{c}{Baselines} \\
\cmidrule(lr){3-4}\cmidrule(lr){5-6}\cmidrule(lr){8-10}
$n$ & $p$ & [1] MaxIE & Num-$h$ & [2] MaxCor & Num-$f^*$ & & [3] Reg$_{Y{\sim}X}$ & [4] Reg$_{A{\sim}X}$ & [5] DM \\
\midrule
\multicolumn{10}{l}{\textit{Objective value (mean\,(sd))}} \\[1pt]
  500 &   20 & $1.07\,(0.15)$ & $1.07\,(0.15)$ & $0.50\,(0.04)$ & $0.50\,(0.04)$ & & \multicolumn{3}{c}{\multirow{2}{*}{\textit{do not directly optimise $h$ or $f^*$}}} \\
 1000 &  100 & $1.09\,(0.12)$ & $1.09\,(0.12)$ & $0.54\,(0.04)$ & $0.54\,(0.04)$ & & \multicolumn{3}{c}{} \\[4pt]
\multicolumn{10}{l}{\textit{Wall-clock time (ms)}} \\[1pt]
  500 &   20 & $0.36$ & $142.8$ & $0.56$ & $488.8$ & & $0.28$ & $0.32$ & $7.5$ \\
 1000 &  100 & $9.28$ & $1{,}951.6$ & $9.36$ & $19{,}277.6$ & & $9.00$ & $8.92$ & $56.5$ \\
 1000 &  500 & $220.3$ & \textit{not run} & $219.7$ & \textit{not run} & & $218.0$ & $218.0$ & $579.6$ \\
   500 & 1000 & $168$ & \multicolumn{7}{c}{\textit{n/a \textemdash{} dual implementation ($p \geq n$)}} \\ 
\bottomrule
\end{tabular}%
}
\end{table}

\subsection{Inference}
\label{sec:inf_sim}

We validate Propositions~\ref{prop:costest} and~\ref{prop:power} and compare
the cosine test against the IUT.
For the top panel of Table~\ref{tab:inference} we use the AR(1)
DGP~\eqref{eq:dgp} with $n=200$ and $p \in \{20, 40, 80\}$
($\rho=0.75$, $\tau=0.25$, $\sigma_\epsilon=0.5$).
For the type~I error rows, $\balpha$ and $\bbeta$ are set to isolate each
null scenario of~\eqref{eq:H0_moments}: (i) both paths null
($\balpha = \bbeta = \mathbf{0}$), (ii) $\beta$-path null
($\balpha = \gamma\mathbf{e}_1$, $\bbeta = \mathbf{0}$), and (iii) $\alpha$-path null
($\balpha = \mathbf{0}$, $\bbeta = \gamma\mathbf{e}_1$), with
magnitude $\gamma=1$, where $\mathbf{e}_j$ denotes the $j$th unit vector.  For the power rows both paths share a single
nonzero entry: $\balpha_j = \bbeta_j = \gamma$ for $j=1$ and zero elsewhere,
with $\gamma \in \{0.10, 0.20\}$.
For the bottom panel (dual, $p=200$, $n\in\{20,40,60,80,100,120\}$) we
use the same structural equations~\eqref{eq:dgp} but with i.i.d.\
$\bX\sim\mathcal{N}(0,I_p)$ and $\tau=0$; the type~I error rows use
a single nonzero entry of magnitude~$1$ on the active path, and the power
rows use $\gamma\in\{0.30,0.50\}$ on a single shared entry
($\balpha_1=\bbeta_1=\gamma$, all others zero), with values chosen to give
discriminating power across the $n$ range shown.
All cells use $1{,}000$ replicates.

Table~\ref{tab:inference} compares empirical rejection rates at nominal
level $\alpha=0.05$.  Under $H_0$, the cosine test holds size near the
nominal level across all three null scenarios and all $p$ values, confirming
Proposition~\ref{prop:costest}.  The IUT is extremely conservative when the
$\beta$-path is null (near-zero rejection), because it requires
both path F-tests to reject simultaneously; this over-conservatism also
manifests when both paths are absent.  Under $H_1$, the cosine test has
consistently higher power, with all entries in the discriminating range
$0.10$--$0.87$ to facilitate comparison.  The advantage over the IUT grows
with both $p$ and weaker signals---consistent with property~(iii) of
Proposition~\ref{prop:power}, which predicts that power concentrates at its
asymptotic limit more slowly for smaller signals relative to the detection
threshold.  At signal $= 0.10$ the cosine test reaches 21\%, 16\%, and
10\% for $p = 20, 40, 80$ respectively, compared to 5\%, 2\%, and 1\%
for the IUT; at signal $= 0.20$ the gap is most striking at $p=80$ (45\% vs.\ 3\%).

\begin{table}[htbp]
\centering
\caption{Empirical rejection rates at nominal $\alpha=0.05$, $1{,}000$ datasets per cell.
\textit{Top panel}: cosine test and IUT, $n=200$, $p\in\{20,40,80\}$, $T\sim t(p-1)$.
\textit{Bottom panel}: dual cosine test only, $p=200$, $n\in\{40,\ldots,140\}$,
$T\sim t(n-2)$; IUT not applicable for $p>n$.
\textit{Type~I error rows}: single nonzero entry (magnitude~1) on the active path.
\textit{Power rows}: both paths share a single nonzero entry of magnitude~$\gamma$.
Bold: higher rejection rate per cell (top panel only).}
\label{tab:inference}
\smallskip
\footnotesize
\setlength{\tabcolsep}{5pt}
\begin{tabular}{ll cc cc cc}
\toprule
\multicolumn{8}{l}{\textit{Top panel: cosine test and IUT, $n=200$}} \\[1pt]
 & & \multicolumn{2}{c}{$p = 20$} & \multicolumn{2}{c}{$p = 40$} & \multicolumn{2}{c}{$p = 80$} \\
\cmidrule(lr){3-4} \cmidrule(lr){5-6} \cmidrule(lr){7-8}
 & & Cosine & IUT & Cosine & IUT & Cosine & IUT \\
\multicolumn{8}{l}{\textit{Type~I error} (nominal 0.05)} \\[1pt]
 & Both paths null           & 0.048 & 0.000 & 0.055 & 0.003 & 0.050 & 0.000 \\
 & $\beta$-path null         & 0.053 & 0.003 & 0.053 & 0.000 & 0.053 & 0.000 \\
 & $\alpha$-path null        & 0.036 & 0.043 & 0.035 & 0.044 & 0.013 & 0.063 \\
\multicolumn{8}{l}{\textit{Power} (both paths active, shared nonzero entry)} \\[1pt]
 & Signal $= 0.10$ & \textbf{0.210} & 0.052 & \textbf{0.159} & 0.016 & \textbf{0.097} & 0.005 \\
 & Signal $= 0.20$ & \textbf{0.840} & 0.662 & \textbf{0.717} & 0.292 & \textbf{0.450} & 0.032 \\
\midrule
\multicolumn{8}{l}{\textit{Bottom panel: dual cosine test, $p=200$, $T\sim t(n-2)$, IUT not applicable}} \\[2pt]
 & & $n=40$ & $n=60$ & $n=80$ & $n=100$ & $n=120$ & $n=140$ \\[1pt]
\multicolumn{8}{l}{\textit{Type~I error} (nominal 0.05)} \\[1pt]
 & Both paths null           & 0.049 & 0.055 & 0.053 & 0.053 & 0.052 & 0.043 \\
 & $\beta$-path null         & 0.029 & 0.026 & 0.020 & 0.010 & 0.007 & 0.004 \\
 & $\alpha$-path null        & 0.048 & 0.057 & 0.052 & 0.061 & 0.061 & 0.055 \\
\multicolumn{8}{l}{\textit{Power} (both paths active, shared nonzero entry)} \\[2pt]
 & Signal $= 0.30$ & 0.078 & 0.107 & 0.156 & 0.178 & 0.194 & 0.233 \\
 & Signal $= 0.50$ & 0.128 & 0.194 & 0.296 & 0.395 & 0.488 & 0.534 \\
\bottomrule
\end{tabular}
\end{table}
The top panel confirms Proposition~\ref{prop:costest} in the $p<n$ regime
($n=200$, $p\in\{20,40,80\}$).
The $\beta$-path-null scenario holds size near nominal across all $p$.
The $\alpha$-path-null scenario is conservative and worsens as $p$ increases
(0.036, 0.035, 0.013 for $p=20,40,80$), confirming the structural asymmetry
of Section~\ref{sec:inference}: the $O(\sqrt{n/p})$ bias from the rank-1
correction to $\SZ$ shrinks only as $n/p\to 0$.
The IUT is severely conservative when both paths are null or the $\beta$-path
is null, because it requires both path F-tests to reject simultaneously.
Under $H_1$, the cosine test has consistently higher power, with the advantage
most pronounced at larger $p$ and weaker signals: at signal $=0.20$ it reaches
84\%, 72\%, and 45\% for $p=20,40,80$, versus 66\%, 29\%, and 3\% for the IUT.

The bottom panel validates Proposition~\ref{prop:dual}(iii) in the $p>n$ regime
($p=200$, $n\in\{40,\ldots,140\}$), exhibiting the same $\alpha$/$\beta$-null
asymmetry with the roles of $n$ and $p$ exchanged.
The both-paths-null and $\alpha$-path-null scenarios hold size near nominal
throughout (0.043--0.061), confirming the $t(n-2)$ approximation.
The $\beta$-path-null scenario is conservative and worsens as $n$ grows
(0.029 at $n=40$ down to 0.004 at $n=140$): in the dual the exact-test
direction is $\beta$-path null, so as $n/p$ increases the test becomes
increasingly conservative on that path---the mirror image of the primal
$\alpha$-path-null conservatism.
Power grows steadily with $n$, consistent with the $\sqrt{n-2}$ scaling of
the noncentrality~\eqref{eq:dual_noncentrality}: at signal $=0.50$, power
rises from $0.13$ at $n=40$ to $0.53$ at $n=140$.

\begin{figure}[tp]
\centering
\includegraphics[width=0.85\linewidth]{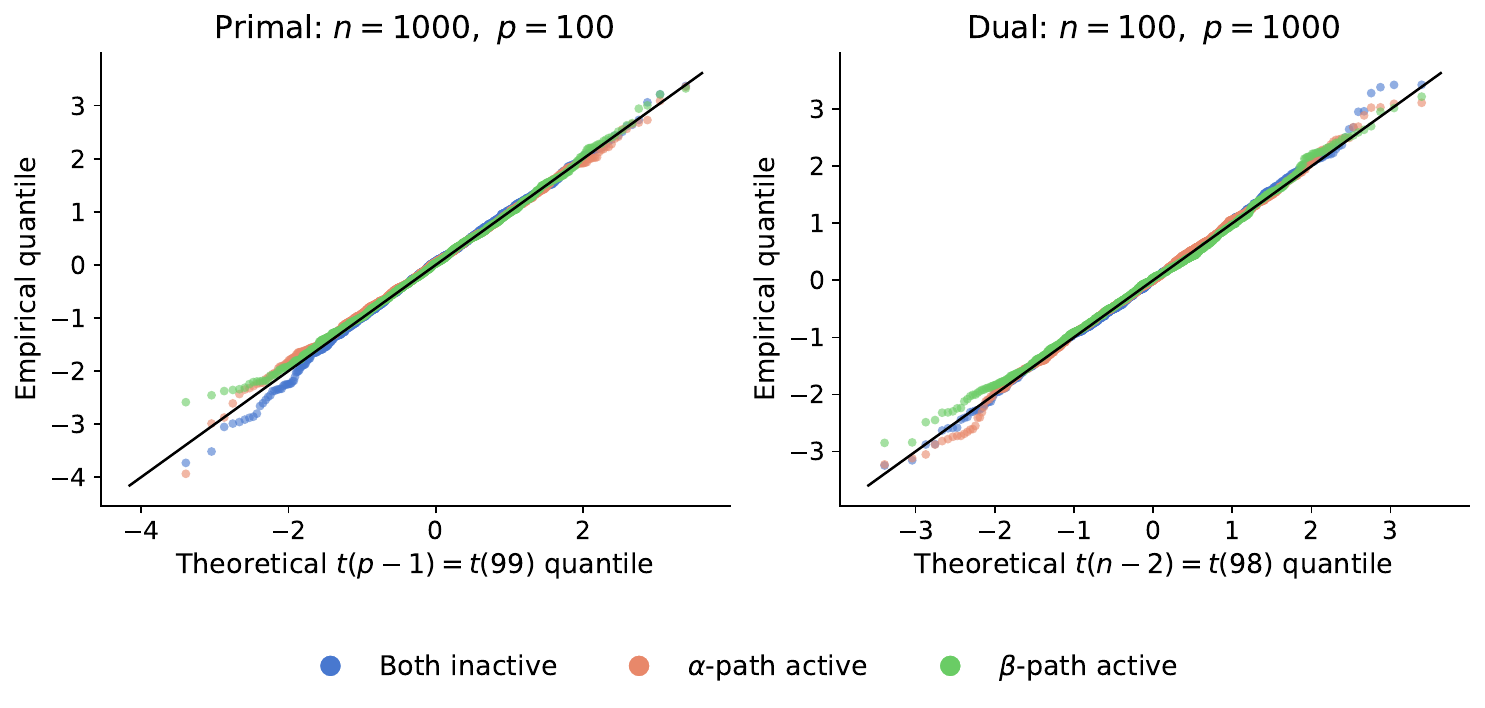}
\caption{QQ plots of the cosine test statistic against $t$ theoretical quantiles
under three null scenarios ($1{,}000$ replicates; DGP as in text).
Null scenarios: \emph{both inactive} ($\balpha=\bbeta=\mathbf{0}$, blue);
\emph{$\beta$-path null} (red); \emph{$\alpha$-path null} (green).
\textit{Left} (primal, $n=1000$, $p=100$): $T$ vs $t(99)$, confirming Proposition~\ref{prop:costest}.
\textit{Right} (dual, $n=100$, $p=1000$): $T$ vs $t(98)$, confirming Proposition~\ref{prop:dual}(iii).}
\label{fig:null_dist}
\end{figure}

Figure~\ref{fig:null_dist} uses the AR(1) DGP~\eqref{eq:dgp}
($\rho=0.75$, $\tau=0$, $\sigma_\epsilon=0.5$) with $1{,}000$ replicates per
null scenario, at primal settings $n=1000$, $p=100$ and dual settings
$n=100$, $p=1000$.
The empirical quantiles of $T$ lie on the diagonal against their theoretical
references under all three null scenarios, confirming both Proposition~\ref{prop:costest}
and Proposition~\ref{prop:dual}(iii).

\begin{figure}[tp]
  \centering
  \includegraphics[width=0.92\textwidth]{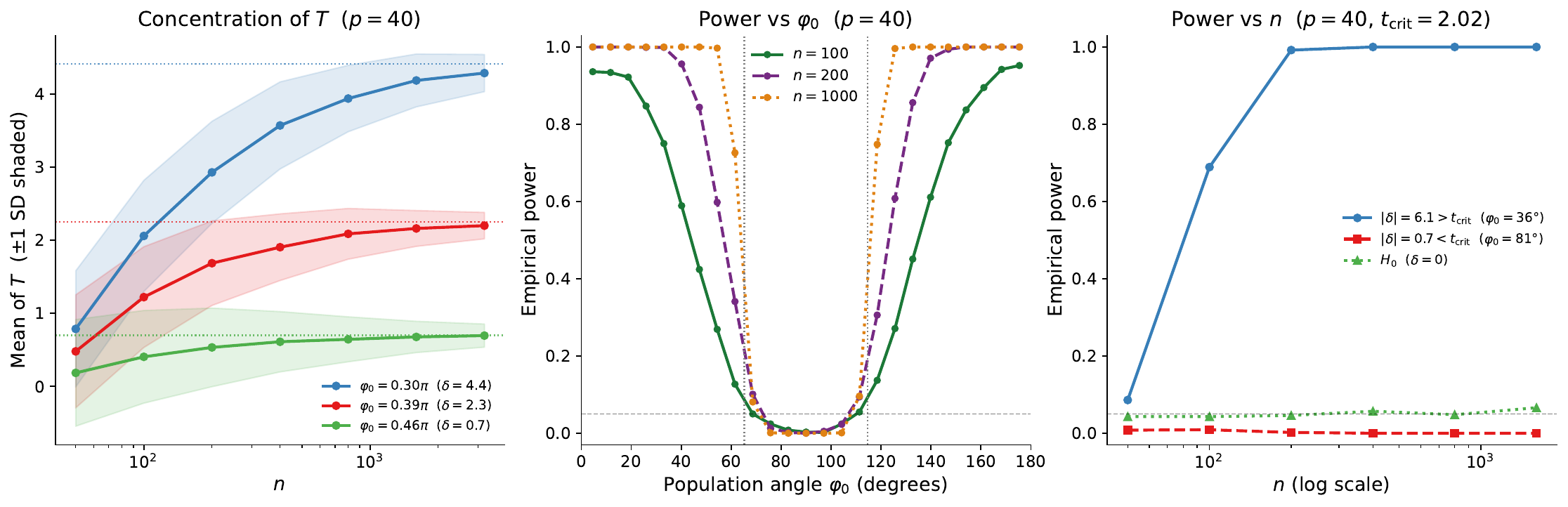}
  \caption{Empirical validation of Proposition~\ref{prop:power}
    ($1{,}000$ simulations per point, $\alpha=0.05$; DGP as in text, $p=40$).
    \textit{Left}: mean$(T)$ and $\pm1$ SD band for
    $\varphi_0\in\{55^{\circ},70^{\circ},84^{\circ}\}$
    ($\delta\in\{3.09,1.61,0.46\}$); dotted horizontals mark theoretical $\delta$;
    band collapses as $n\to\infty$, confirming property~(i).
    \textit{Centre}: empirical power vs $\varphi_0$ for
    $n\in\{100,200,1000\}$; grey verticals mark the detection threshold,
    confirming property~(ii).
    \textit{Right}: empirical power vs $n$ for $\varphi_0=55^{\circ}$
    (above threshold), $84^{\circ}$ (below), and $90^{\circ}$ (null),
    confirming property~(iii).}
  \label{fig:power}
\end{figure}

Figure~\ref{fig:power} uses a two-entry DGP with $\Sigma=I_p$, $\tau=0$,
$\sigma_\epsilon=0.5$: $\balpha = s\cdot\mathbf{e}_1$ and
$\bbeta = s(\cos\varphi_0\,\mathbf{e}_1 + \sin\varphi_0\,\mathbf{e}_2)$ with $s=0.5$,
so $\varphi_0$ is the $\ell_2$ angle between $\balpha$ and $\bbeta$.
The noncentrality $\delta = \cot\tilde\varphi_0\cdot\sqrt{p-1}$ is computed
analytically from the population $\SZ$ at the true path vectors
($\tilde\varphi_0$ is the $\SZ^{-1}$-metric angle).
The left panel uses $p=40$, $\ell_2$ angles
$\varphi_0\in\{55^\circ,70^\circ,84^\circ\}$
($\delta\in\{3.09,1.61,0.46\}$),
$n\in\{50,100,200,400,800,1600,3200\}$;
the centre panel sweeps $\varphi_0\in(0^\circ,180^\circ)$ at $p=40$,
$n\in\{100,200,1000\}$; and the right panel sweeps $n\in\{50,\ldots,1600\}$
at $p=40$ for one angle above and below the detection threshold
($1{,}000$ replicates per point, $\alpha=0.05$).

\begin{figure}[tp]
  \centering
  \includegraphics[width=0.92\textwidth]{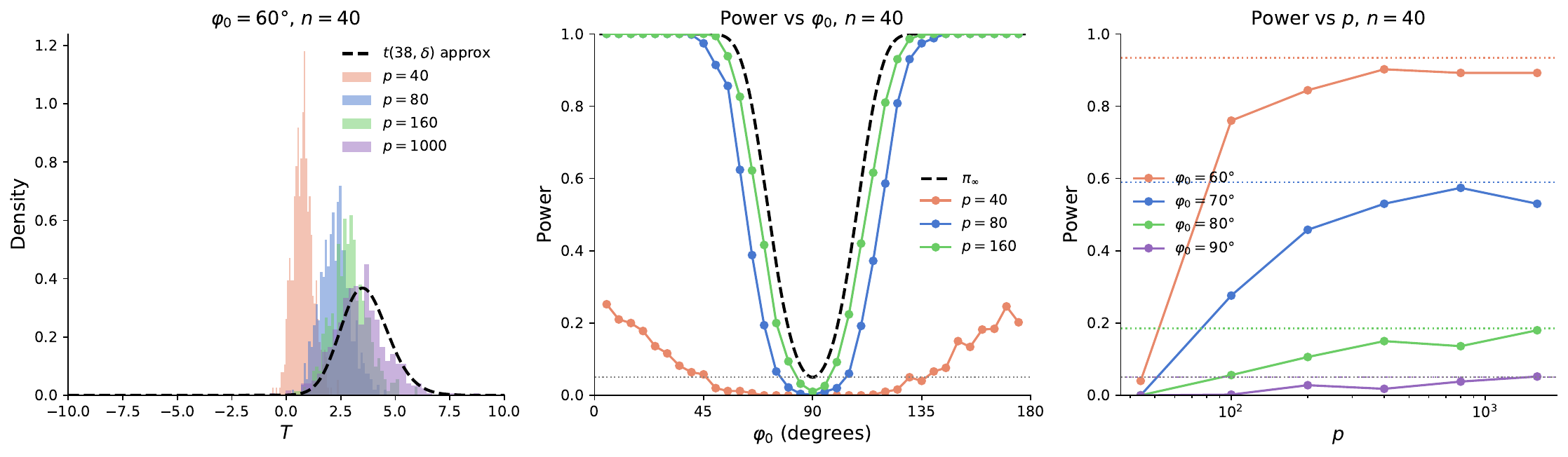}
  \caption{Empirical validation of Proposition~\ref{prop:dual}(iii)
    ($1{,}000$ replicates per point, $n=40$, $\alpha=0.05$; DGP as in text).
    \textit{Left}: empirical density of $T$ at $\varphi_0=60^{\circ}$ for
    $p\in\{40,80,160,1000\}$; dashed curve is $t(38,\delta)$ with
    $\delta=\cot 60^{\circ}\cdot\sqrt{38}\approx 3.56$;
    density converges to the approximation as $p$ grows.
    \textit{Centre}: empirical power vs $\varphi_0$ for $p\in\{40,80,160\}$,
    showing the U-shape; dashed curve is $\pi_\infty$~\eqref{eq:pi_inf}.
    \textit{Right}: empirical power vs $p$ for
    $\varphi_0\in\{60^{\circ},70^{\circ},80^{\circ},90^{\circ}\}$;
    dotted horizontals mark $\pi_\infty\in\{0.93,0.59,0.19,0.05\}$,
    confirming saturation with no sharp detection threshold.}
  \label{fig:power_dual}
\end{figure}

Figure~\ref{fig:power_dual} supports Proposition~\ref{prop:dual}(iii).
In the dual regime, $\cos\varphi$ converges to $\cos\phi_0(\bA,\bY)$~\eqref{eq:cosphi_limit},
so $T$ has a non-degenerate limiting distribution rather than a point mass.
We use a dense-signal DGP with $n=40$, $\tau=0$: $\balpha$ a random unit
vector in $\mathbb{R}^p$, $\bbeta=\cos\varphi_0\,\balpha+\sin\varphi_0\,\bv_\perp$
with $\bv_\perp\perp\balpha$, $X_{ij}=\mathrm{SNR}\cdot\sqrt{p}\cdot\alpha_j A_i+\varepsilon_{ij}$,
$Y_i=\mathrm{SNR}\cdot\sqrt{p}\cdot\bX_i^\top\bbeta+\varepsilon_i$, $\mathrm{SNR}=0.5$,
all noise i.i.d.\ $\mathcal{N}(0,1)$ ($1{,}000$ replicates per $p$, $\alpha=0.05$);
the $\sqrt{p}$ rescaling keeps total signal amplitude constant so that $\varphi_0$ remains fixed as $p$ grows.

\section{Application: Proteomic Mediators of Age on Dementia}
\label{sec:application}

\paragraph{Motivation: what should a proteomic aging composite do?}
Proteomic aging clocks regress chronological age on protein abundances and
use the resulting age gap to stratify disease risk~\citep{oh2023,argentieri2024,tian2025}.
Such clocks track biological aging well but, as we show below, add limited
independent disease information beyond chronological age.
Conversely, a protein score trained directly to predict disease captures
dementia-associated proteins regardless of whether they are driven by the
aging process---including reverse-causal signals from preclinical disease.
A composite useful for \emph{both} monitoring biological aging and predicting
disease must do two things simultaneously: track age-driven proteomic change
(so it responds when biological aging is slowed) and predict the disease
endpoint (so that response is clinically meaningful).
The MaxIE objective $\hat\alpha(\bw)\hat\beta(\bw)$ encodes both requirements
directly: $\hat\alpha(\bw)$ is age tracking and $\hat\beta(\bw)$ is disease
relevance, and maximising their product finds the unique composite jointly
optimal on both.

\paragraph{Data and methods.}
We apply the method to a UK Biobank proteomics dataset ($n = 38{,}383$,
$p = 2{,}916$ proteins~\citep{tian2025}; treatment $\bA$: chronological age;
outcome $\bY$: binary dementia status, $1{,}205$ cases, $3.1\%$ prevalence).
The dataset is split 70:30 into training/test sets; the linear model is a
working approximation justified by low prevalence and large
$n$~\citep{vanderweele2015}.
We compare \textbf{MaxIE} against \textbf{Lasso-Age} ($\ell_1$-penalised
age-on-proteins clock) and \textbf{Lasso-Y} ($\ell_1$-penalised
dementia-on-proteins predictor), both cross-validated.
All statistics are on the held-out test set.

\paragraph{MaxIE achieves the highest indirect effect, in and out of sample.}
The cosine test yields $\cos\hat\varphi = 0.107$ ($p = 6.4 \times 10^{-9}$), strongly rejecting the global null.
MaxIE attains the highest indirect effect on both training ($h = 0.0142$) and
test ($h = 0.0058$) sets, confirming that directly optimising $\hat\alpha\hat\beta$
yields a strictly better composite than optimising either path alone.
Lasso-Age reaches only $h = 0.0018$ on the test set ($3.2\times$ lower) despite
its high $r_{MA}$, because its composite is nearly orthogonal to dementia after
age adjustment ($r_{M\perp Y} = 0.031$).
Lasso-Y achieves $h = 0.0031$ ($1.9\times$ lower than MaxIE) because its low
age tracking ($r_{MA} = 0.534$) limits the $\alpha$-path.
MaxCor leads on $f^*$ ($= 0.1313$ test) as expected; MaxIE is competitive
($f^* = 0.1199$) while retaining the advantage on the indirect effect.
Rankings are consistent across splits, confirming out-of-sample generalisation.
MaxIE and Lasso-Y exhibit suppression ($P_M > 1$); the mediation index $f^*$
provides a stable comparison in this regime~\citep{mackinnon2000,vanderweele2015}.

\begin{figure}[htbp]
\vspace{-4pt}
\centering
\includegraphics[width=0.96\linewidth]{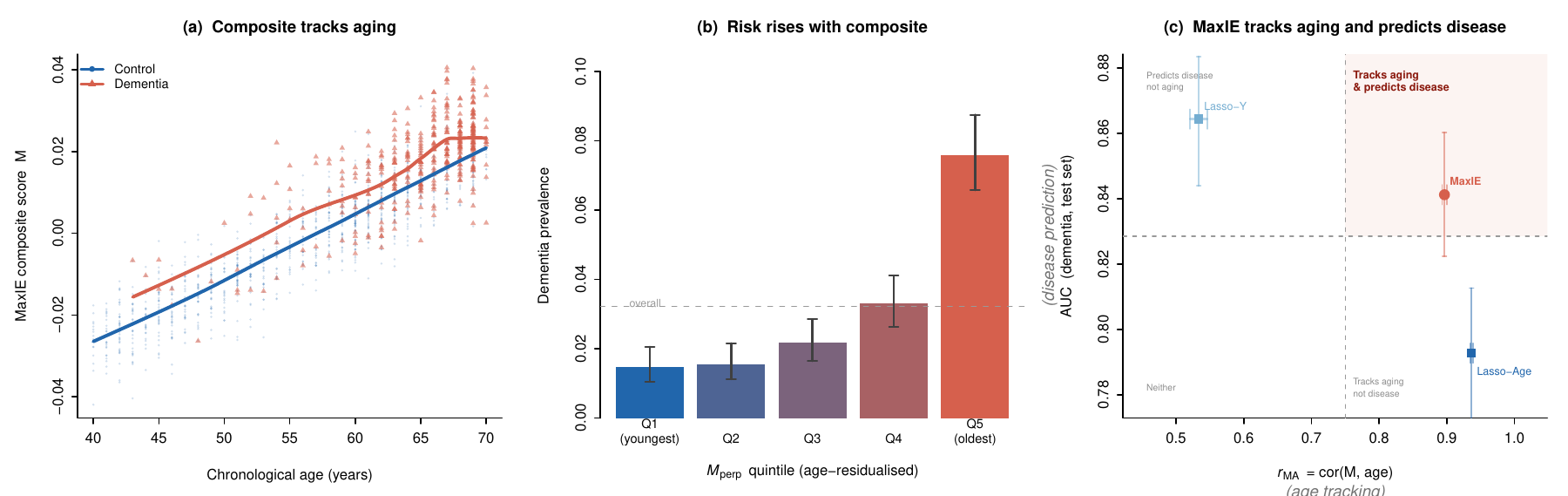}
\caption{Validation of the MaxIE composite mediator on the UK Biobank test
set ($n = 11{,}515$; 371 dementia cases).
\textbf{(a)} MaxIE composite score $M$ versus chronological age, with loess
smoothers for controls (blue) and dementia cases (red).
The composite rises steeply with age and is systematically elevated in
cases, consistent with accelerated biological aging.
\textbf{(b)} Dementia prevalence by quintile of $M_\perp = \bQ_A(\bX\hat\bw^*)$,
the composite residualised on chronological age, with 95\% Wilson confidence
intervals. Prevalence rises monotonically from $1.4\%$ (Q1) to $7.6\%$ (Q5),
a five-fold gradient independent of age.
\textbf{(c)} Each method as a point in the space of age tracking
($r_{MA} = \cor(M, A)$, x-axis) and disease prediction
(AUC for dementia, y-axis).
Error bars are 95\% bootstrap confidence intervals ($B = 1{,}000$).
The upper-right quadrant (shaded) is the region of jointly high age tracking
and disease prediction; MaxIE uniquely occupies this quadrant while each
baseline excels on only one dimension.}
\label{fig:application}
\end{figure}

\paragraph{MaxIE jointly optimises age tracking and disease prediction; the baselines do not.}
Figure~\ref{fig:application}(a) shows that the MaxIE composite rises
steeply with age and is systematically elevated in dementia cases relative
to controls, consistent with accelerated biological aging in cases.
Figure~\ref{fig:application}(b) shows dementia prevalence rising monotonically
across quintiles of $M_\perp = \bQ_A(\bX\hat\bw^*)$ (age-residualised composite),
from $1.4\%$ in the lowest quintile to $7.6\%$ in the highest---a five-fold
gradient independent of chronological age.
Figure~\ref{fig:application}(c) places all three methods in the space of
age tracking ($r_{MA}$, x-axis) and disease prediction (AUC, y-axis).
Lasso-Age occupies the bottom-right: high age tracking ($r_{MA} = 0.937$)
but low disease prediction (AUC $= 0.793$).
Lasso-Y occupies the top-left: high disease prediction (AUC $= 0.864$)
but low age tracking ($r_{MA} = 0.534$), reflecting dementia-specific
rather than aging-driven proteomic variation.
MaxIE uniquely occupies the top-right quadrant ($r_{MA} = 0.897$,
AUC $= 0.841$), outperforming each baseline on the dimension the baseline
neglects.

\begin{table}[htbp]
\centering
\caption{Mediation statistics for composite protein scores applied to the
UK Biobank proteomics dataset (treatment $\bA$: chronological age; outcome
$\bY$: dementia status). MaxIE optimises $\hat\alpha\hat\beta$ using all $p=2{,}916$ proteins;
Lasso-Age and Lasso-Y are single-criterion penalised regression baselines.
MaxCor is included for completeness.
Top panel: the two proposed objectives, bold = best per row per split.
Bottom panel: mediation summary statistics.}
\label{tab:ukb}
\smallskip
\resizebox{\linewidth}{!}{%
\setlength{\tabcolsep}{5pt}
\begin{tabular}{ll *{4}{r} *{4}{r}}
\toprule
& & \multicolumn{4}{c}{Training set} & \multicolumn{4}{c}{Test set} \\
\cmidrule(lr){3-6} \cmidrule(lr){7-10}
Metric & Sym.
  & \multicolumn{1}{c}{MaxIE}
  & \multicolumn{1}{c}{MaxCor}
  & \multicolumn{1}{c}{Lasso-Age}
  & \multicolumn{1}{c}{Lasso-Y}
  & \multicolumn{1}{c}{MaxIE}
  & \multicolumn{1}{c}{MaxCor}
  & \multicolumn{1}{c}{Lasso-Age}
  & \multicolumn{1}{c}{Lasso-Y} \\
\midrule
\multicolumn{10}{l}{\textit{Proposed objectives}} \\[2pt]
Indirect effect & $h\,(\hat\alpha\hat\beta)$ & \textbf{0.0142} & 0.0121 & 0.0028 & 0.0042 & \textbf{0.0058} & 0.0049 & 0.0018 & 0.0031 \\
Mediation index & $f^*$ & 0.2790 & \textbf{0.3131} & 0.0427 & 0.1681 & 0.1199 & \textbf{0.1313} & 0.0289 & 0.1220 \\
\midrule
\multicolumn{10}{l}{\textit{Mediation summary}} \\[2pt]
cor$(M,A)$         & $r_{MA}$                      & 0.9130 & 0.8433 & 0.9487 & 0.5401 & 0.8966 & 0.8213 & 0.9366 & 0.5335 \\
cor$(M_\perp,Y)$   & $r_{M\perp Y}$                & 0.3056 & 0.3713 & 0.0450 & 0.3112 & 0.1337 & 0.1598 & 0.0308 & 0.2287 \\
Total effect       & $\hat\tau$                    & 0.0035 & 0.0035 & 0.0035 & 0.0035 & 0.0037 & 0.0037 & 0.0037 & 0.0037 \\
Prop.\ mediated    & $\hat\alpha\hat\beta/\hat\tau$ & 4.073  & 3.470  & 0.803  & 1.189  & 1.570  & 1.334  & 0.477  & 0.837  \\
\bottomrule
\end{tabular}%
}
\end{table}

\section{Conclusion}
\label{sec:conclusion}

Finding a composite biomarker that maximally transmits the effect of an
exposure through a high-dimensional mediator set is a natural goal in
molecular epidemiology, yet it has not previously been pursued as a direct
optimisation problem.  We showed that the indirect effect $\hat\alpha\hat\beta$,
despite being non-convex in the weight vector $\bw$, admits a closed-form
global optimum: the solution lies in a two-dimensional subspace and is
recovered by two linear solves followed by a single $\arctan$ evaluation.
The same geometric argument yields an asymptotic test for the global
mediation null: under $H_0$, the path-angle cosine has a $\mathrm{Beta}(1/2,(p-1)/2)$
distribution on $[-1,1]$ after squaring, and the reparametrised statistic
$T = \cos\varphi\sqrt{(p-1)/(1-\cos^2\!\varphi)} \sim t(p-1)$
asymptotically---a free by-product of the solver that yields a two-sided test
for any composite mediation and directional one-sided tests for concordant
and suppression mediation.
A companion algorithm, MaxCor, directly maximises the bounded mediation index
$f^*$, providing a scale-free summary that remains interpretable in the
suppression regime where proportion mediated is undefined.

Simulations confirm that both algorithms outperform single-arm OLS baselines
by wide margins and that each leads on its own criterion, while the cosine
test controls size asymptotically and dominates the intersection-union test in power.
In the UK Biobank proteomics application, MaxIE jointly optimises both
dimensions of a useful biological age measure: it tracks chronological age
($r_{MA} = 0.897$) comparably to a dedicated aging clock while achieving
higher dementia discrimination (AUC $= 0.841$) than either single-criterion
baseline.
The conventional aging clock (Lasso-Age) fails on disease prediction;
the disease predictor (Lasso-Y) fails on age tracking.
MaxIE is the unique composite that does both, providing a proteomic biological
age measure that is simultaneously anchored to the aging process and
predictive of disease outcome~\citep{oh2025,argentieri2024}.

A practical advantage of the proposed approach is its compatibility with
federated data infrastructures.  Large-scale omics studies are increasingly
conducted across multiple cohorts or biobanks, where data-sharing agreements,
privacy regulations, or logistical constraints prevent centralising
individual-level records.  Because MaxIE and MaxCor depend on the data only
through cross-product summaries $(\ba, \bz, \SZ, \|\bA\|^2, \|\bY\|^2)$,
each site can contribute these quantities locally and the coordinator runs the
solver once on the pooled summaries---no individual records ever leave the
site.  The cosine test requires the same inputs, so estimation and inference
are both fully federated.
\paragraph{Limitations.}
The Gaussian error assumption in Propositions~\ref{prop:costest}
and~\ref{prop:power} is used to establish uniformity of the whitened
path direction; robustness to non-Gaussian errors remains an open question.
The dual cosine test ($t(n-2)$) is derived as $p\to\infty$ with $n$ fixed;
its behaviour under proportional growth $p/n \to c > 1$ requires random
matrix methods and is open, as is formal size control for the primal test
in the proportional regime $p/n \to c \in (0,1)$.
Uncertainty quantification beyond the global null test is also limited:
a confidence interval for $\hat\alpha(\hat\bw^*)$, $\hat\beta(\hat\bw^*)$,
or $h(\hat\bw^*)$ that conditions on the estimated $\hat\bw^*$ does not
correct for selection bias, in the same way that an in-sample $R^2$ overstates
out-of-sample fit.
The honest approach is sample splitting, as in the UK Biobank analysis;
a bias-corrected interval for the population indirect effect $h(\bw^*_0)$
requires separate treatment, for instance via bootstrap procedures that
mimic the full estimation pipeline.
Finally, the weight vector $\bw^*$ identifies the optimal composite direction but does not support inference about individual mediators: a nonzero $w^*_j$ may reflect that $X_j$ is correlated with the true mediating subspace rather than that $X_j$ has an independent mediation effect.
Variable selection based on $|w^*_j|$ is reasonable for constructing or interpreting the composite, but identifying which features are individually causal mediators requires a separate inferential procedure, such as marginal mediation tests or a sparse penalised extension.
\paragraph{Future directions.}
A natural sparse extension would replace the two linear solves with
$\ell_1$-penalised regressions; the bisector geometry carries over exactly,
but the null distribution of $\cos\varphi$ under penalisation requires
separate theoretical treatment.
The dual formulation opens a direct path to nonlinear extensions: replacing
the linear kernel $\bK = \bX\bX^\top$ with an RBF or Mat\'{e}rn kernel
gives a kernelised MaxIE that optimises the indirect effect in a
reproducing kernel Hilbert space, with the bisector geometry preserved in
the feature space.
A natural and important direction is relaxing Assumption A3: replacing the
linear structural model with a semiparametric model for the NIE
(in the spirit of \citep{tchetgen2012}) would yield a more robust
estimator of the optimal composite mediator, though deriving the analogous
closed form under nonparametric path models is non-trivial.
Further extensions include the multi-outcome setting where $\bY$ is a matrix,
longitudinal exposures, and cross-study aggregation under federated constraints.


\section*{Data Availability Statement}
The UK Biobank data used in Section~\ref{sec:application} are available to
approved researchers through the UK Biobank access procedure
(\url{https://www.ukbiobank.ac.uk/enable-your-research/apply-for-access}).
The \texttt{optmed} Python and R packages implementing MaxIE, MaxCor, and the
cosine test, together with all simulation scripts required to reproduce the
tables and figures, are available at
\url{https://github.com/statzihuai/optmed}.

\section*{Acknowledgements}
Z.H.\ was supported by NIH/NIA awards AG089509, AG066206, and AG066515.

\section*{Conflict of Interest}
The author declares no conflict of interest.

\section*{AI Use Disclosure}
Claude (developed by Anthropic) was used in the preparation of this
manuscript to support mathematical exposition, proof writing, simulation
code, and drafting of results and conclusions. All core ideas, theoretical
contributions, and scientific claims originate with the author. All
mathematical results, proofs, and numerical findings were independently
verified by the author.

\end{document}